\documentclass[journal=jpccck,manuscript=article]{achemso}
\usepackage[version=3]{mhchem} 
\usepackage[utf8]{inputenc}
\usepackage{array}
\usepackage{wrapfig}
\usepackage{multirow}
\usepackage{tabularx}

\author{Raphael M. Tromer}
\affiliation[IF UnB]
{University of Brasília, Institute of Physics, Brasília, Federal District, Brazil.}

\author{Bruno Ipaves}
\affiliation[State University of Campinas]
{Department of Applied Physics and Center for Computational Engineering and Sciences, State University of Campinas, Campinas, São Paulo, Brazil.}

\author{Marcelo L. Pereira Junior}
\affiliation[ENE-UnB]{Department of Electrical Engineering, College of Technology, University of Brasília, Brasília, Federal District 70910-900, Brazil.}
\alsoaffiliation[RiceUni]{Department of Materials Science and NanoEngineering, Rice University, Houston, Texas 77005, United States.}

\author{Cristiano F. Woellner}
\affiliation[Federal University of Parana]
{Physics Department, Federal University of Parana, UFPR, Curitiba, PR, 81531-980, Brazil.}
\alsoaffiliation[CICTI]
{Interdisciplinary Center for Science, Technology, and Innovation (CICTI),  Federal University of Parana,  UFPR, Curitiba, PR, 81530-000, Brazil.}

\author{Kun Cai}
\affiliation[Harbin Institute of Technology]
{School of Science, Harbin Institute of Technology, Shenzhen 518055, China.}

\author{Douglas S. Galvao}
\email{galvao@ifi.unicamp.br}
\affiliation[State University of Campinas]
{Department of Applied Physics and Center for Computational Engineering and Sciences, State University of Campinas, Campinas, São Paulo, Brazil.}
\title{On the Electronic, Mechanical and Optical Properties of Superhard Cross-Linked Carbon Nanotubes (Tubulanes)}

\keywords{Carbon, Tubulane, DFT, porous materials}
\usepackage{xcolor}
\begin{document}


\begin{abstract}

We have investigated the electronic, optical, and mechanical properties of six structures belonging to the Tubulanes-cross-linked carbon nanotube family. Our results highlight the remarkable anisotropic mechanical behavior of these materials, distinguishing them from isotropic structures, such as diamond. Notably, the 8-tetra-22 structure has a higher Young's modulus ($Y_M$) along the $z$-direction compared to diamond. Unlike diamonds, the mechanical properties of Tubulanes are direction-dependent, with significant variations in Young's Modulus (2.3 times). Additionally, the Poisson's ratio is highly anisotropic, with at least one direction exhibiting an approximately zero value. The inherent anisotropy of these materials enables tunable mechanical properties that depend on the direction of applied stress. Regarding their electronic properties, all Tubulane structures studied possess indirect electronic band gaps, dominated by $2p$ orbitals. The band dispersion is relatively high, with band gaps ranging from 0.46 eV to 2.74 eV, all of which are smaller than that of diamond. Notably, the 16-tetra-22 structure exhibits the smallest bandgap (0.46 eV), making it particularly interesting for electronic applications. Additionally, these structures exhibit porosity, which provides an advantage over denser materials, such as diamond. Considering the recent advances in the synthesis of 3D carbon-based materials, the synthesis of tubulane-like structures is within our present-day technological capabilities. 

\end{abstract}

\section{Introduction}

The search for materials with exceptional mechanical, electronic, and optical properties has generated significant interest in carbon-based nanostructures, such as carbon nanotubes (CNTs) and graphene. Among these, the cross-linked carbon nanotube structures, known as tubulanes, represent a class of materials with remarkable strength, stability, and electronic properties. Notably, tubulanes exhibit super hardness and structural integrity under extreme conditions, making them attractive candidates for a wide range of applications, from impact-resistant materials to next-generation electronics \cite{baughman1993tubulanes, rysaeva2019elastic}.

Several studies have demonstrated that specific arrangements of CNTs can significantly impact the mechanical and electronic properties of the resulting structure. For example, high-pressure CNT networks exhibit enhanced mechanical strength and resilience due to the interlinked nature of the tubes \cite{sajadi20193d, baughman1993tubulanes, rysaeva2019elastic}. The electronic properties of tubulanes, like those of other carbon nanomaterials, are highly dependent on their atomic configuration. Similar to other carbon allotropes, tubulanes can exhibit either metallic or semiconducting behavior depending on their topology \cite{ajayan1992smallest, ebbesen1992large, iijima_1991, sajadi20193d}.

Theoretical studies have played a crucial role in predicting the behavior of these complex nanostructures. Density functional theory (DFT) has proven to be a reliable method for investigating various properties, including the electronic and optical characteristics of carbon-based materials, offering valuable insights into their electronic band gaps \cite{tromer2021dft, ipaves2024tuning, rysaeva2019elastic}. In addition, molecular dynamics simulations are frequently employed to explore thermal stability and mechanical responses under varying conditions \cite{baughman1993tubulanes, sajadi20193d, rysaeva2019elastic}. The combination of these simulation techniques has significantly improved our ability to predict and tailor the properties of tubulanes and other nanostructures \cite{rysaeva2019elastic}.

A promising aspect of tubulanes is their potential for novel applications in diverse fields. The combination of mechanical strength and electronic versatility in these diamond-like superhard phases could play a key role in areas such as energy storage, sensing, and UV collectors \cite{baughman2002carbon, rysaeva2019elastic, tromer2021dft}. Recent studies have also highlighted their potential use in ballistic applications \cite{sajadi20193d}. These emerging applications underline the need for further investigation into their optical properties and structural stability under diverse environmental conditions.

In this work, we have investigated the structural, mechanical, electronic, and optical properties of tubulanes using DFT and \textit{ab initio} molecular dynamics (AIMD) simulations. Building upon prior theoretical studies, we have examined the stability of two families of tubulanes: tetragonal and hexagonal. The tetragonal structures, named 8-tetra-22, 8-tetra-33, and 16-tetra-22, and the hexagonal structures, named 12-hexa-33, 24-hexa-20, and 36-hexa-33, were comprehensively analyzed. We assess their mechanical behavior under uniaxial compression and investigate their electronic and optical properties, providing new insights into their potential for ultraviolet-blocking applications.

\section{Methodology}

We have carried out DFT simulations using the CASTEP \cite{Castep_2005} and SIESTA \cite{Soler_2002} codes. CASTEP is a plane wave basis set method. We have used CASTEP for the mechanical and electronic analyses of the tubulanes. The calculations were performed using the generalized gradient approximation (GGA) and the Perdew-Burke-Ernzerhof (PBE) \cite{Perdew1996} functional to represent the exchange-correlation term. To improve the description of non-covalent interactions in the GGA-PBE calculations, we used a semi-empirical dispersion correction scheme of Tkatchenko and Scheffler (PBE+D) \cite{DFT_vdW}. To replace the core electrons in each atomic species, we adopted norm-conserving pseudopotentials \cite{norm_cons}. We used the Monkhorst-Pack scheme to represent the k-points of integration in the Brillouin zone, with k-points set to $5\times 5\times 10$ and an energy cutoff of $300$ eV \cite{Monkhorst_1976}. The optimization process was carried out using the Broyden, Fletcher, Goldfarb, and Shanno (BFGS) scheme \cite{bfgs}. Both atoms and lattice parameters were optimized simultaneously. We considered the forces in each atom to be equal to $0.01$ eV/\r{A} as the convergence criterion. We adopted an energy variation smaller than $0.5 \times 10^{-6}$ eV/atom for each self-consistent field (SCF) cycle as a convergence criterion. The elastic properties were extracted using the geometrically optimized structures. We have also investigated the structural stability of the tubulane structures at room temperature by carrying \textit{ab initio} molecular dynamics (AIMD) simulations using an NPT ensemble, with a time step of 1 fs. The unit cell was replicated 2 times along each direction.  

For the simulations of optical properties, we have used the SIESTA software \cite{Soler_2002}, a DFT-based code that employs basis sets of localized atomic orbitals. We have used the functional GGA/PBE, which is equivalent to that used in CASTEP simulations. We have used the same geometries obtained with the CASTEP and verified that SIESTA and CASTEP produce practically identical electronic bandgap values.

We used the Troullier-Martins pseudopotentials in the SIESTA simulations to describe core electrons with a basis set double-zeta plus polarization functions (DZP). The cutoff value for kinetic energy is 250 Ry, and the k-points are $12\times12\times12$, using the Monkhorst-Pack scheme.

The optical simulations were modeled by a system immersed in the region of the externally applied electrical field, operating in the linear regime associated with electromagnetic radiation. In the linear regime, the typical value of the external electrical field, which does not cause significant distortion in the structures, is $1.0$ V/\r{A} \cite{Fadaie2016}. We have considered polarization along one specific direction separately ($x$, $y$, or $z$). 

The starting point consists of obtaining the real part $\epsilon_1$ and imaginary part $\epsilon_2$ of the dielectric constant function. These quantities are extracted directly from the Kramers-Kronig relation and Fermi's golden rule, respectively.

All optical coefficients are calculated by relating the real and imaginary parts of the dielectric constant function. The absorption coefficient $\alpha$, the reflectivity  $R$, and the refractive index $\eta$ are given by:
\begin{equation}
    \alpha (\omega )=\sqrt{2}\omega\bigg[(\epsilon_1^2(\omega)+\epsilon_2^2(\omega))^{1/2}-\epsilon_1(\omega)\bigg ]^{1/2},
\end{equation}
\begin{equation}
    R(\omega)=\bigg [\frac{(\epsilon_1(\omega)+i\epsilon_2(\omega))^{1/2}-1}{(\epsilon_1(\omega)+i\epsilon_2(\omega))^{1/2}+1}\bigg ]^2 \text{, and}
\end{equation}
\begin{equation}
    \eta(\omega)= \frac{1}{\sqrt{2}} \bigg [(\epsilon_1^2(\omega)+\epsilon_2^2(\omega))^{1/2}+\epsilon_1(\omega)\bigg ]^{2}.
\end{equation}

The accuracy of the optical coefficients is related to a good description of the electronic bandgap value. In other words, the more accurate the bandgap value is, the more precisely the optical transitions will be described. It is well known that GGA/PBE does not accurately reproduce the electronic band gap values of semiconductor materials. This approximation tends to underestimate the experimental value obtained with more robust methods \cite{Johnson_1998}. Other approximations, such as the HSE06 hybrid functional, yield more accurate bandgap values \cite{Kishore2017}. 
Here, we used the HSE06 functional implemented in the Gaussian16 software to obtain the electronic bandgap energy of carbon tubulanes with higher precision. We used the parameters that produce a good description of the bandgap value of cubic diamond as a reference.

Then, we use the scissor operator available in SIESTA to correct the optical transitions using the electronic band gap values extracted from Gaussian16. 
In SIESTA, this correction is defined by a quantity called the scissor operator, which produces a shift in the unoccupied states, given by:

\begin{equation}
\mathrm{scissor}=E_{gap}^{\mathrm{HSE06}}-E_{gap}^{\mathrm{PBE}}.
\label{scissor}
\end{equation}

In the literature, some works have employed this approximation to correct the optical transitions generated with the scissor operator, yielding results similar to those of more sophisticated methods, such as GW ones \cite{Nayebi2016, Ljungberg2017}. 

\section{Results}

\subsection{Structural Stability}

\begin{figure}[htb!]   
    \begin{center}  
        \includegraphics[width=0.4\linewidth]{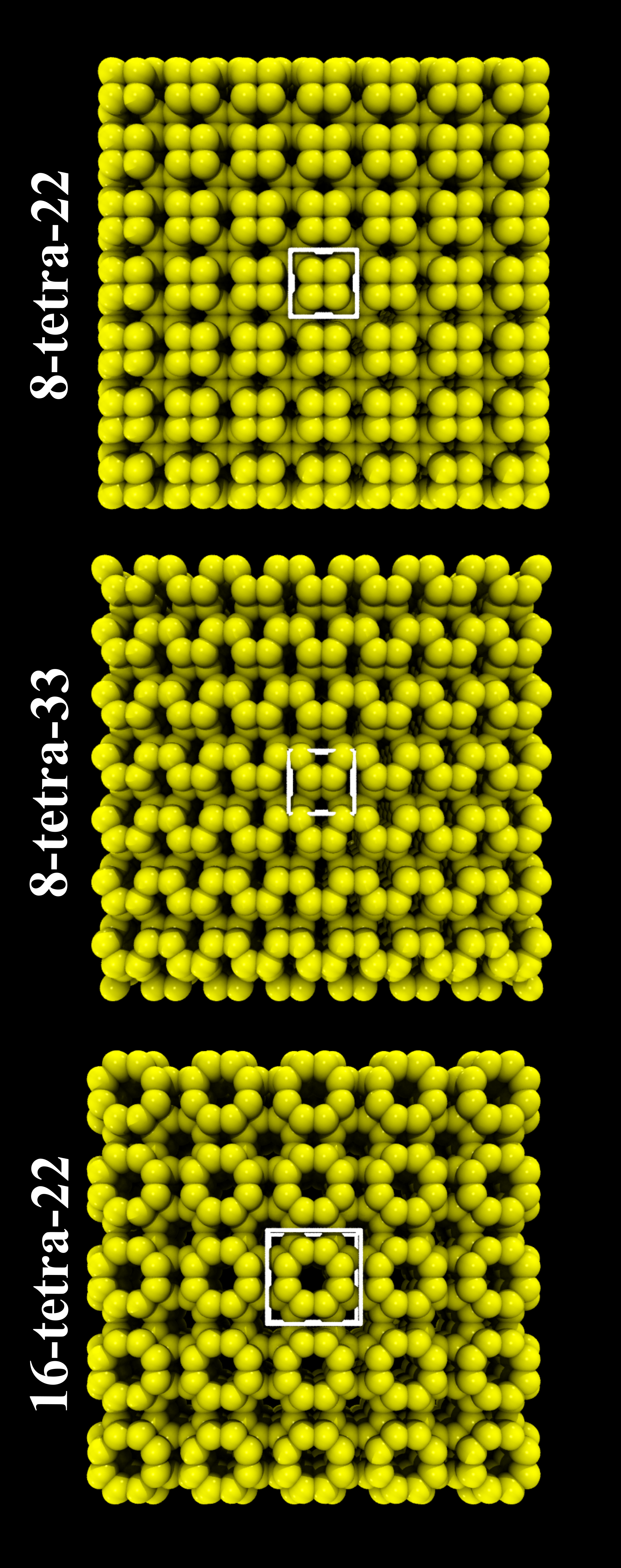}
        \caption{Schematic representation of the tetragonal family structures: 8-tetra-22, 8-tetra-33, and 16-tetra-22. The unit cell in white was replicated $2\times 2\times 2$.}
        \label{fig:structure1}
    \end{center}
\end{figure}

\begin{figure}[htb!]
\begin{center}
\includegraphics[width=0.4\linewidth]{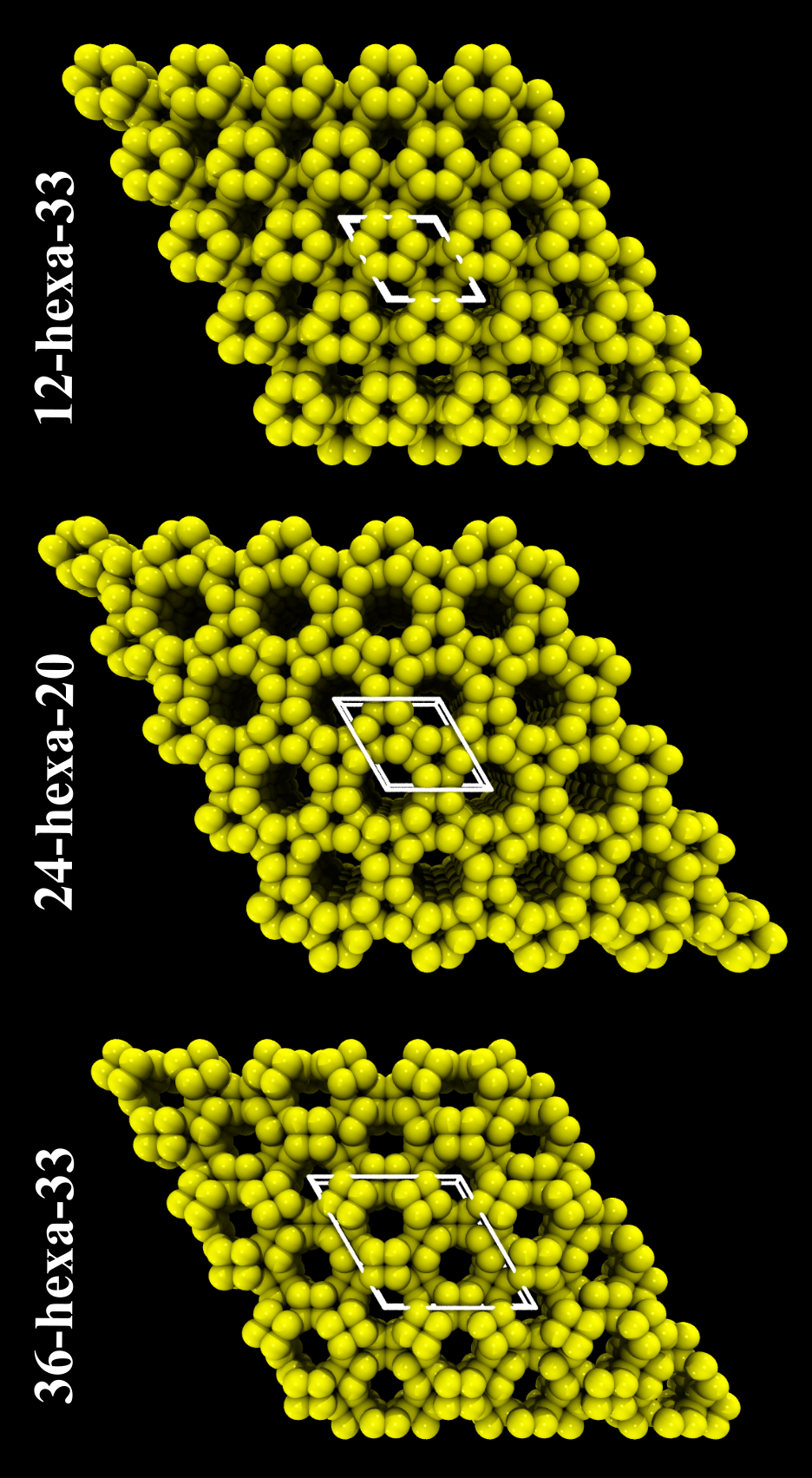}
\caption{Schematic representation of the hexagonal family structures: 12-hexa-33, 24-hexa-20, and 36-hexa-33. The unit cell in yellow was replicated $2\times 2\times 2$.}
\label{fig:structure2}
\end{center}
\end{figure}

We present in Figure \ref{fig:structure1} the tetragonal family, with the optimized structures 8-tetra-22, 8-tetra-33, and 16-tetra-22. The unit cell is marked in white and replicated $2\times 2\times 2$ for building the crystal. Figure \ref{fig:structure2} illustrates the optimized structures for the hexagonal family called supercell 12-hexa-33, 24-hexa-20, and 36-hexa-33, respectively. Viewing of other material plans is available in the supplementary material.

Table \ref{tab:formation} displays the structural parameters for the optimized tubulane structures calculated with CASTEP. We included the values for the cubic diamond structure for comparison. From Table \ref{tab:formation} we observe that the diamond structure is isotropic, while the 8-tetra-33 phase is orthotropic. The other structures are transversely isotropic, exhibiting identical properties along the $x$ and $y$ directions but distinct behavior along the $z$ direction. Note that 16-tetra-22 presents a density value close to that of diamond, $3.48$ and $3.51$ g/cm$^3$, respectively. The lowest density value is $2.904$ g/cm$^3$ of the 24-hexa-20 structure.

\begin{table}
    \centering
    \begin{tabular}{ | m{2.2cm} | m{1.2cm}| m{1.4cm} | m{3.4cm} | m{3.2cm}| m{1.5cm} | m{1.0cm} |} 
    \hline
    \hline
    Structure  & Atoms & Tubule type of largest channel & Space group name and 
    number & Unit cell optimized lattice parameters [\AA] & Density [g/cm$^3$]  \\
    \hline
    8-tetra-22  & 8 & (2,2) & $I4/mmm$ (139) & $a=4.358$ $b=4.358$ $c=2.497$ & 3.425\\
    \hline
    8-tetra-33  & 8 & (3,3) & $Imma$ (74) & $a=4.922$ $b=2.533$ $c=4.181$ & 3.179\\
    \hline
    16-tetra-22 & 16 & (4,4) & $I4/mmm$ (139) & $a=6.539$ $b=6.539$ $c=2.506$ & 3.048\\
    \hline
    12-hexa-33  & 12 & (3,3) & $P6_{3}/mmc$ (194) & $a=6.090$ $b=6.090$ $c=2.538$ & 2.984\\
    \hline
    24-hexa-20 & 24 & (6,0) & $P6/mcc$ (192) & $a=6.808$ $b=6.808$ $c=4.174$ & 2.904\\
    \hline
    36-hexa-33 & 36 & (3,3) & $R\overline{3}m$ (166) & $a=10.451$ $b=10.451$ $c=2.484$ & 3.140\\
    \hline
     Diamond & 8 & - & $Fd\overline{3}m$ (227) & $a=3.579$ $b=3.579$ $c=3.579$ & 3.510 \\
     \hline
    \hline
    \end{tabular}
    \caption{Summary of the calculated structural parameters for all Tubulane structures.}
    \label{tab:formation}
\end{table}

We have also investigated the structural stability of the tubulanes at 300 K. We carried out \textit{ab initio} molecular dynamics (AIMD) simulations of 2 ps runs with an NPT ensemble, using a time step of 1 fs. In these simulations, the unit cell of each structure was replicated $2 \times 2\times 2$ times. In Figures \ref{fig:SI-5-hexa-MD} and \ref{fig:SI-6-hexa-MD} in supplementary materials, we present representative AIMD snapshots for the tubulane structures, showing that tubulanes are structurally stable at room temperature (300~K). These results are significant because an approach was used to synthesize similar structures at ambient conditions \cite{Zhong1379}. Therefore, such techniques could potentially be used to create the tubulane structures.

\subsection{Mechanical properties}

To determine the role of tubulane's unusual topologies in their mechanical behavior, we calculated the elastic constants using first-principles calculations provided by CASTEP. As discussed in the Methods section, the CASTEP Elastic Constants module provides the full 6$\times$6 tensor of elastic constants, from which we can calculate the bulk modulus ($B$), Young Modulus ($Y_M$), and Poisson coefficients shown in Table \ref{tab:mec_prop}.

The mechanical properties of the Tubulane structures exhibit significant anisotropy, unlike the isotropic nature of diamond, as shown in Table \ref{tab:mec_prop}. The Young’s modulus varies considerably depending on the direction, with some structures showing higher stiffness along the $z$ direction than along the $x$ or $y$ directions. Notably, the 8-tetra-22 structure has a Young’s modulus of 1195.35 GPa along the z-direction, exceeding that of diamond (1046.31 GPa). These results highlight the potential of Tubulanes for applications requiring high stiffness along specific axes. Additionally, the bulk modulus of 8-tetra-22 (396.28 GPa) approaches that of diamond (435.03 GPa), making it the stiffest among the Tubulane structures investigated here.

The results presented in Table \ref{tab:mec_prop} also reveal the highly anisotropic nature of Poisson’s ratio in Tubulanes, in stark contrast to the uniform value of 0.1 observed in diamond. This directional dependence suggests that Tubulanes exhibit distinct deformation behaviors along different axes. Some structures exhibit extremely low Poisson’s ratios (close to zero) along specific directions, such as 24-hexa-20 and 16-tetra-22, indicating minimal lateral contraction under axial compression. Furthermore, 8-tetra-33 shows auxetic behavior (negative Poisson’s ratio) along the $yz$ plane, which is rare in conventional materials. This auxetic property can be helpful in mechanically responsive materials that expand under tension.

\begin{table}
    \centering
    \begin{tabular}{|c|c|c|c|c|c|c|c|}
    \hline
    Structure    & $K$ & $(Y_M)_x$ & $(Y_M)_y$ & $(Y_M)_z$  & $\nu_{xy} (\nu_{yx})$ & $\nu_{zx} (\nu_{xz})$ & $\nu_{yz} (\nu_{zy})$\\
    \hline
    8-tetra-22  & 396.28 & 900.62 & 900.62 & 1195.35 & \vtop{\hbox{\strut 0.172}\hbox{\strut (0.172)}} & \vtop{\hbox{\strut 0.046}\hbox{\strut (0.034)}} & \vtop{\hbox{\strut 0.0344}\hbox{\strut (0.046)}}\\
    \hline
    8-tetra-33  & 351.94 & 391.17 & 908.17 & 574.04 & \vtop{\hbox{\strut 0.113}\hbox{\strut (0.263)}} & \vtop{\hbox{\strut 0.621}\hbox{\strut (0.423)}} & \vtop{\hbox{\strut -0.084}\hbox{\strut (-0.053)}}\\
    \hline
    16-tetra-22 & 335.94 & 751.56 & 751.56 & 967.89 & \vtop{\hbox{\strut 0.250}\hbox{\strut (0.250)}} & \vtop{\hbox{\strut 0.013}\hbox{\strut (0.010)}}  & \vtop{\hbox{\strut 0.010}\hbox{\strut (0.013)}}\\
    \hline
    12-hexa-33  & 335.20 & 722.71 & 722.71 & 905.14 & \vtop{\hbox{\strut 0.227}\hbox{\strut (0.227)}}  & \vtop{\hbox{\strut 0.059}\hbox{\strut (0.047)}} & \vtop{\hbox{\strut 0.047}\hbox{\strut (0.059)}}\\
    \hline
    24-hexa-20 & 322.65 & 718.60 & 718.60 & 1098.43 & \vtop{\hbox{\strut 0.198}\hbox{\strut (0.198)}}  & \vtop{\hbox{\strut 0.012}\hbox{\strut (0.008)}}  & \vtop{\hbox{\strut 0.008}\hbox{\strut (0.012)}} \\
    \hline
    36-hexa-33 & 340.49 & 689.92 & 689.92 & 1084.43 & \vtop{\hbox{\strut 0.271}\hbox{\strut (0.271)}}  & \vtop{\hbox{\strut 0.027}\hbox{\strut (0.017)}} & \vtop{\hbox{\strut 0.017}\hbox{\strut (0.027)}}  \\
    \hline
     Diamond & 435.03 & 1046.31 & 1046.31 & 1046.31 & 0.100 & 0.100 & 0.100\\
     \hline
    \end{tabular}
    \caption{Summary of calculated mechanical properties for all Tubulane structures. The Young's modulus ($(Y_M)_k$) and bulk modulus ($K$) are given in GPa.}
    \label{tab:mec_prop}
\end{table}

\section{Electronic Properties}

\begin{figure}[htb]
\begin{center}
\includegraphics[width=0.6\linewidth]{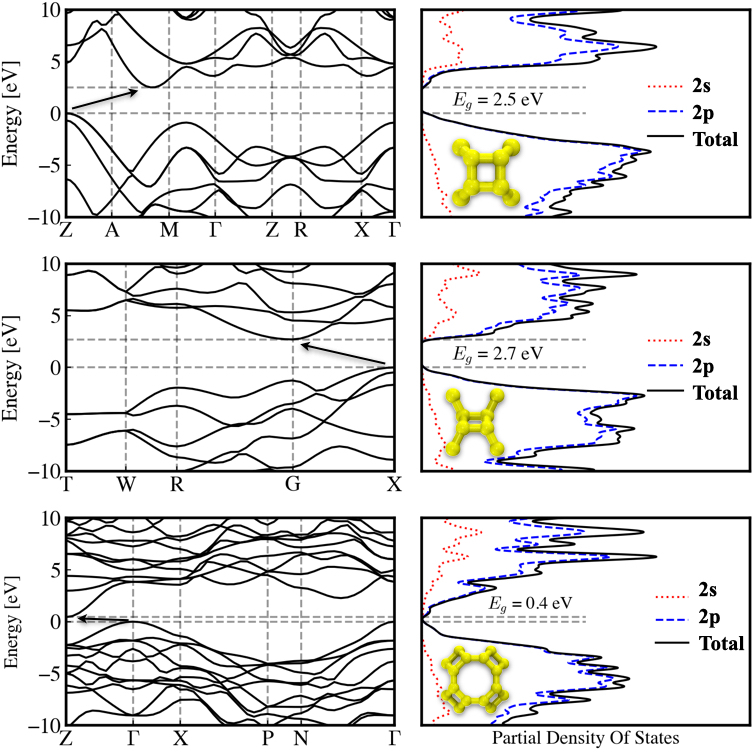}
\caption{Electronic band structure of the tetragonal optimized geometries shown in Figure \ref{fig:structure1}, i.e, 8-tetra22, 8-tetra-33, and 16-tetra-22.
\label{fig:bands_tet}}
\end{center}
\end{figure}

In Figure \ref{fig:bands_tet}, we present the calculated electronic band structure and the corresponding projected density of states (PDOS) of valence atomic orbitals $2s$ and $2p$ for the tetragonal optimized geometries shown in Figure \ref{fig:structure1}. The unit cell is highlighted in yellow.

We note that tetragonal structures exhibit indirect electronic band gap values of 2.5, 2.7, and 0.4 eV for 8-tetra-22, 8-tetra-33, and 16-tetra-22, respectively. The arrow indicates the direct/indirect bandgap trend. Intriguingly, 8-tetra-22 and 16-tetra-22 structures differ only in the tube type chiralities (2,2) and (4,4), respectively. Nevertheless, the bandgap values are significantly different, with 8-tetra-22 exhibiting a larger bandgap value, while 16-tetra-22 has a relatively small one. This occurs because the tubulanes structures are built similarly, where nanotubes are pressed at high pressure until they react to obtain the equilibrium structure. The 16-tetra-22 is the only structure that we observe clearly with tubes compressed, while other structures present different forms, see Figures \ref{fig:SI-1-tetra-UC}, \ref{fig:SI-2-hexa-UC}, \ref{fig:SI-3-tetra}, and \ref{fig:SI-4-hexa} in the supplementary materials. This indicates that the deformations in structure 16-tetra-22 are less significant than in the other structures. As the 16-tetra-22 structure is metallic and carbon nanotubes under strain tend to open the gap, it explains why this structure has a smaller gap than the other structures, where deformations are more evident.

We observe from the PDOS that $2p$ atomic orbitals make the most significant contributions to the tetragonal tubulanes states. The $2s$ atomic orbitals will hybridize with the $2p$ orbitals to form $sp^2$ hybridization. However, this hybridization is minimal near the valence and conduction bands, as the $2s$ valence states have minimal contributions in these regions.

\begin{figure}[htb]
\begin{center}
\includegraphics[width=0.7\linewidth]{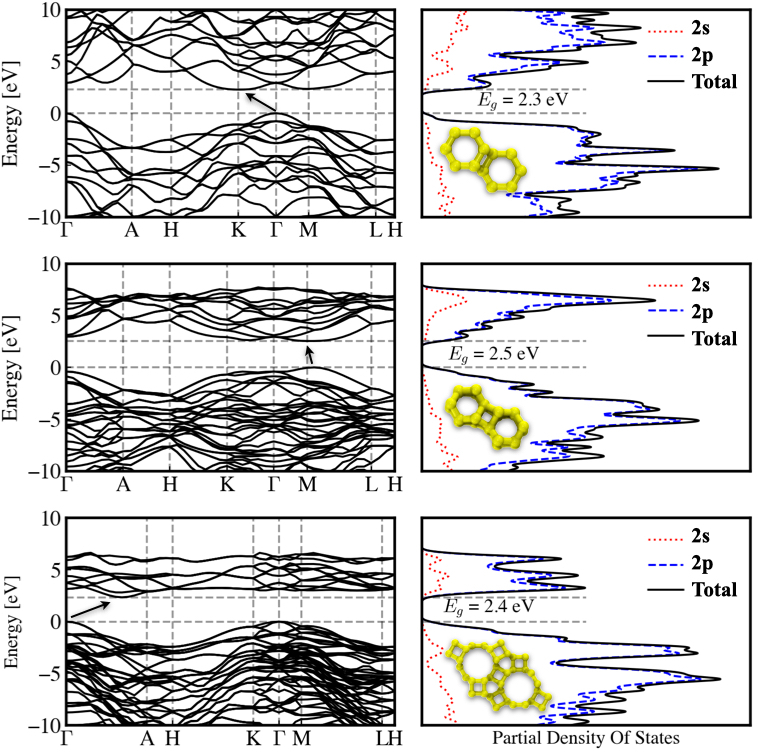}
\caption{Electronic band structure of the hexagonal optimized geometries shown in Figure \ref{fig:structure2}
\label{fig:bands_hexa}, i.e, 12-hexa-33, 24-hexa-20, and 36-hexa-33.}
\end{center}                                                         
\end{figure}

In Figure \ref{fig:bands_hexa}, we present the electronic band structure and the corresponding PDOS for the hexagonal optimized geometries shown in Figure \ref{fig:structure2}. For the hexagonal structures, the band gap values follow the same trend as in the tetragonal case with indirect band gaps and similar values of 2.3, 2.5, and 2.4 eV for 12-hexa-33, 24-hexa-20, and 36-hexa-33, respectively. We also observe from the PDOS that hybridization occurs mainly between the $2s$ and $2p$ valence atomic orbitals.

\begin{table}
    \centering
    \begin{tabular}{|c|c|c|c|}
    \hline
    \hline
    Structure  & E$_{\rm gap,CASTEP(PBE)}$ (PBE/DFT-D) & E$_{\rm gap,Gaussian(HSE06)}$  \\
    \hline
    8-tetra-22  & 2.52 (2.49) & 3.76\\
    \hline
    8-tetra-33  & 2.74 (2.70) & 3.81\\
    \hline
    16-tetra-22 & 0.46 (0.44) & 1.53\\
    \hline
    12-hexa-33  & 2.30  (2.28) & 4.24\\
    \hline
    24-hexa-20 & 2.61  (2.55) & 3.83\\
    \hline
    36-hexa-33 & 2.45 (2.35) & 3.83\\
    \hline
     Diamond & 4.15 (4.19) & 5.40\\
     \hline
    \hline
    \end{tabular}
    \caption{Summary of the calculated electronic properties for all Tubulane structures. All energy values are in eV.}
    \label{tab:electronic}
\end{table}

\subsection{Optical properties}

In the methodology section, we mentioned that the optical simulations were performed with a correction for optical transition implemented in SIESTA, known as the SCISSOR operator. The bandgap value from Gaussian software was used to implement the correction in SIESTA.

\begin{figure}[htb]
\begin{center}
\includegraphics[width=1.0\linewidth]{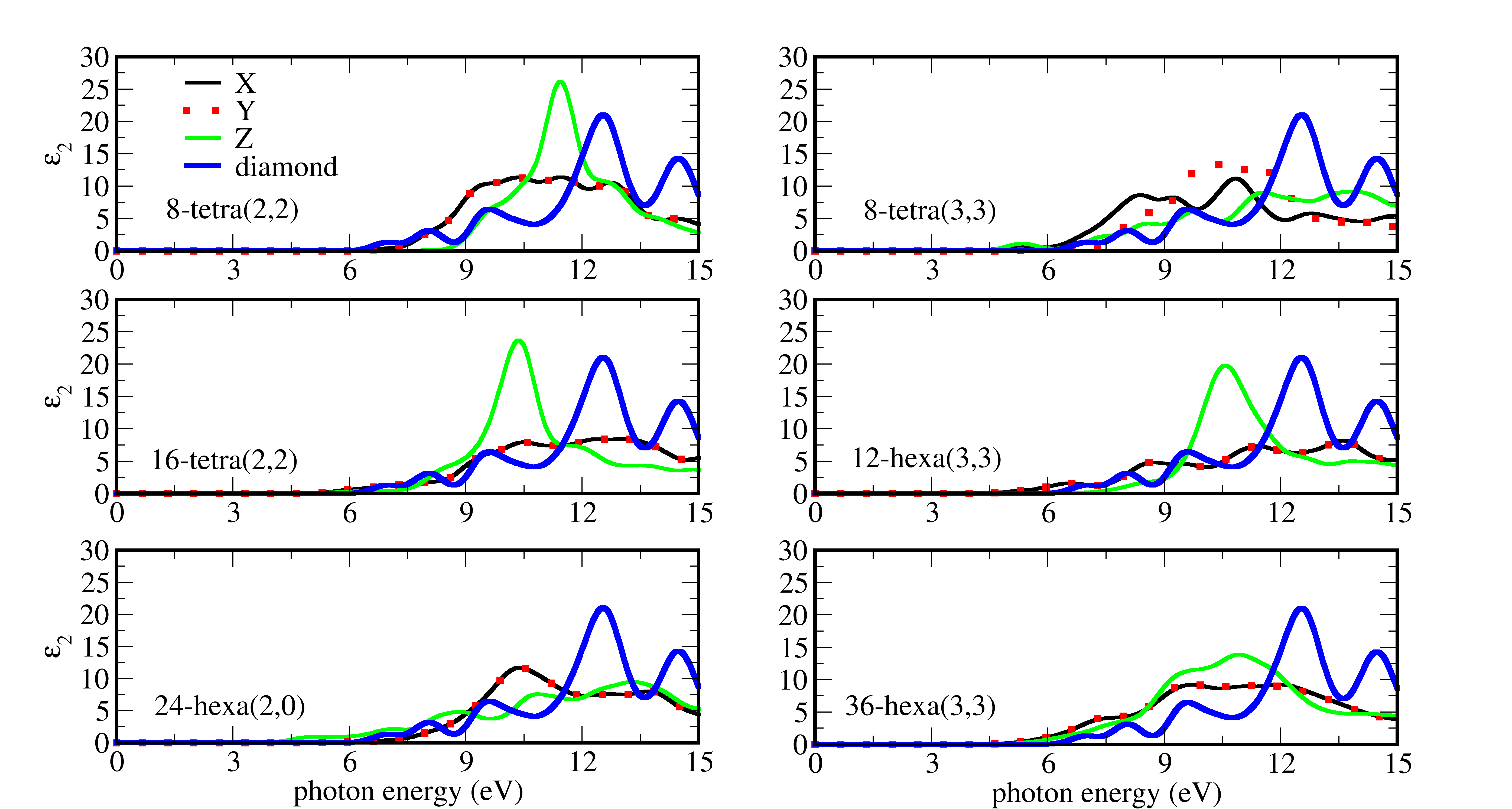}\\
\caption{Imaginary part of the dielectric function \textit{versus} photon energy, for an externally applied electric field polarized along the $x$, $y$, and $z$ directions.}
\label{fig:epsilon_im}
\end{center}
\end{figure}

In Figure \ref{fig:epsilon_im}, we present the imaginary part of the dielectric function as a function of the photon energy for tubulanes and diamond structures.
For each structure, the first peak in the imaginary part, $\epsilon_2$, starts for photon energy values close to the electronic bandgap values. In Table \ref{tab:electronic}, we can see that, except for 16-tetra-22, the structures have wide band gap values. This agrees with a high photon energy value at which the imaginary part of the dielectric constant starts. Although the structure 16-tetra-22 presents a small band gap value, it exhibits an optical transition in the violet range.
Diamond is isotropic along $x$, $y$, and $z$ directions, as expected. In contrast, only the tubulane 8-tetra-33 is anisotropic, while the others are isotropic only along the $x$ and $y$ directions and anisotropic along the $z$ direction. 

Figure \ref{fig:eps} in the supplementary material shows the real part of the dielectric function as a function of photon energy. From these curves, we extracted the static dielectric constant for each structure, taking the value in the limit of high frequencies, or equivalently, for photon energy at zero. For diamond, the value is $5.4$, which is in agreement with the experimental value \cite{epvalue}. For tubulanes, the static dielectric constant is very close to the diamond value, and the maximum difference is smaller than 0.5.

\begin{figure}[htb]
\begin{center}
\includegraphics[width=1.0\linewidth]{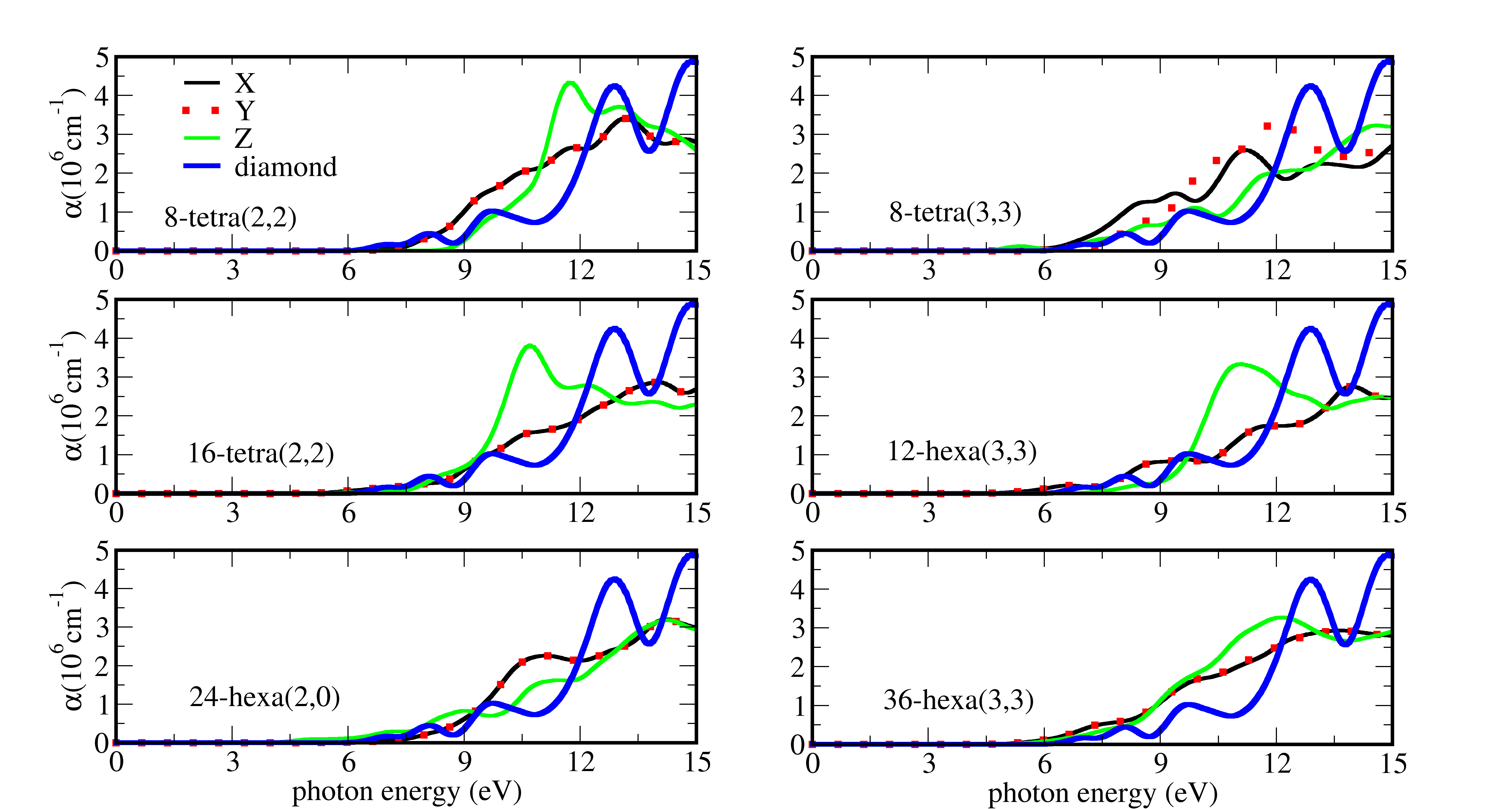}\\
\caption{Absorption coefficient as a function of the photon energy, for an externally applied electric field polarized along the $x$, $y$, or $ Z$ direction. \label{fig:abs}}
\end{center}
\end{figure}

In Figure \ref{fig:abs}, we present the absorption coefficient as a function of photon energy. We observe that for tubulanes, the trends are very similar to diamond, and the peak of maximum absorption intensity is very close. The absorption starts in the ultraviolet range, as expected for systems with large bandgap values. For tubulanes, we observe that the maximum intensity for the absorption coefficient occurs when the external electrical field is polarized along the $z$ direction.

In Figure \ref{fig:refrac}, in the supplementary materials, we present the refractive index as a function of photon energy for the diamond and tubulanes. The behavior of tubulanes is very similar to diamond, in which only the peak for maximum intensity is located at different photon energy values. The refractive index for each structure is obtained when the photon energy is equal to zero, corresponding to high frequencies. The diamond has a refractive index of 2.33, which is close to the experimental value of 2.4 \cite{refrvalue}. For tubulanes, the values are very close to diamond, with a maximum difference of only 0.2.

\begin{figure}
\begin{center}
\includegraphics[width=1.0\linewidth]{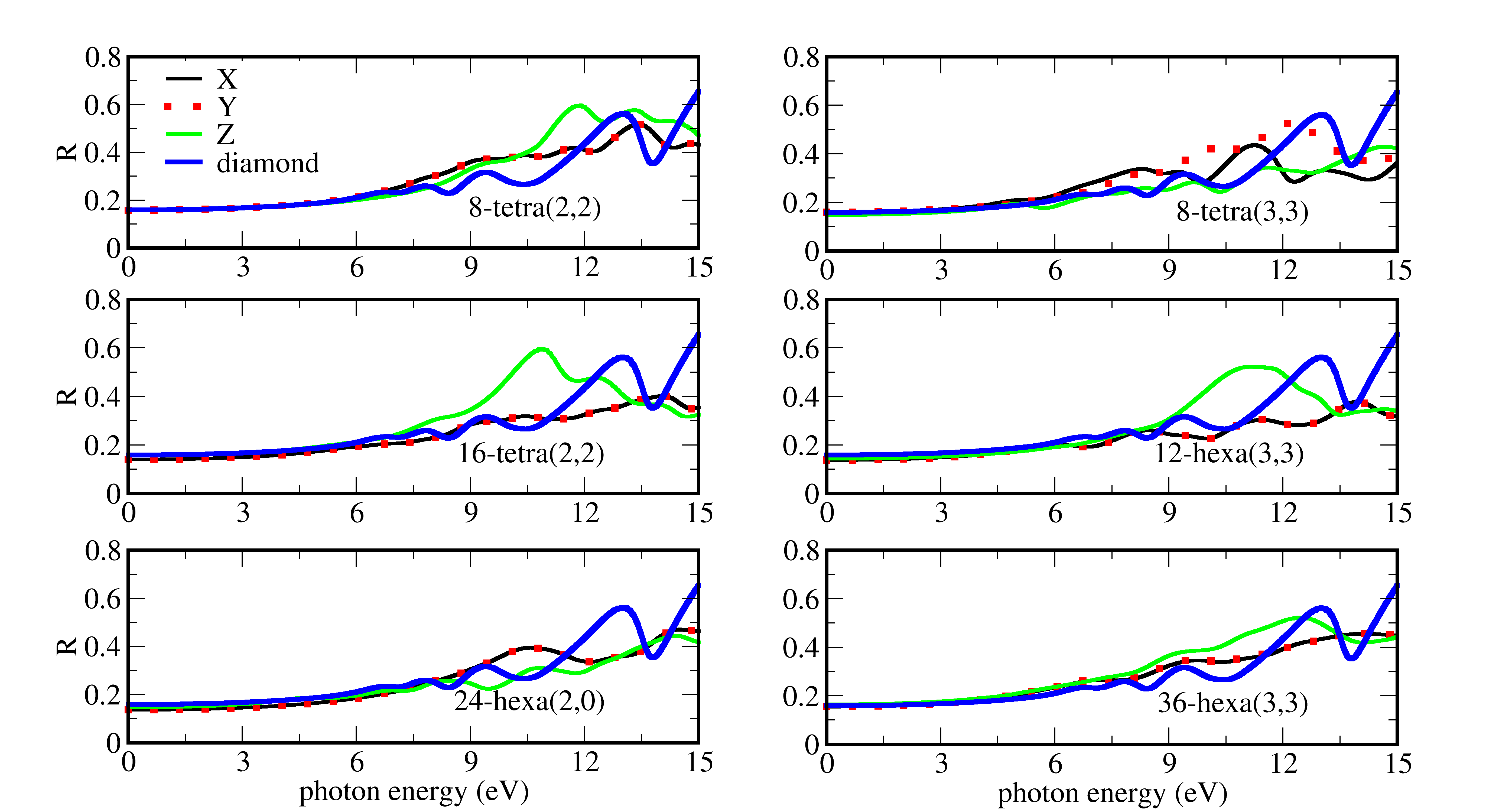}\\
\caption{Reflectivity as a function of the photon energy, for an externally applied electric field polarized along the $x$, $y$, or $ Z$ directions. \label{fig:refl}}
\end{center}
\end{figure}

In Figure \ref{fig:refl}, we have the reflectivity as a function of photon energy for tubulanes and diamond structures. The curves are very similar and exhibit the smallest values in the infrared region, increasing until they reach their highest value in the ultraviolet region. 
For the ultraviolet region, where absorption is maximum for photon energies close to 14-15 eV, the reflectivity is also maximum, and almost 70\% of the incident light will be reflected. This is one indication that tubulanes can be used in ultraviolet block devices.

\section{Conclusions}

Our study highlights the unique combination of anisotropic mechanical, electronic, and optical properties exhibited by Tubulanes-cross-linked carbon nanotube structures. Unlike diamond, these materials display direction-dependent mechanical behavior, with some structures surpassing diamond’s bulk and Young’s modulus along specific directions. The ability to tune mechanical properties by selecting the compression direction makes them particularly attractive for applications requiring high-strength yet adaptable materials. Additionally, the anisotropic Poisson’s ratio, with near-zero values in certain directions and auxetic behavior in specific cases, further expands their potential for tailored mechanical applications.  

From an electronic perspective, all studied Tubulanes exhibit indirect electronic band gaps, with values smaller than that of diamond. Furthermore, the intrinsic porosity of these structures offers advantages over denser materials, such as diamond, enabling applications where a balance between mechanical resilience, lightweight properties, and permeability is required.  

Overall, the unique combination of high mechanical strength, tunable anisotropy, electronic versatility, and porosity positions Tubulanes-cross-linked carbon nanotube structures as promising materials for next-generation nanodevices, impact-resistant coatings, and advanced electronic applications. Their tunable nature makes them highly adaptable for application-specific optimizations, paving the way for innovative material design strategies. Considering the recent advances in the synthesis of 3D carbon-based carbons, the synthesis of tubulane-like structures is within our present-day technological capabilities.
\section*{Acknowledgements}

In memory of Professor Ray H. Baughman, to whom the authors dedicate this work, in recognition of his inspiring scientific contributions and lasting influence.

This work was partially financed by the Coordination for the Improvement of Higher Education Personnel (CAPES), the Brazilian National Council for Scientific and Technological Development (CNPq), and the São Paulo Research Foundation (FAPESP). The authors thank the Center for Computational Engineering \& Sciences (CCES) at Unicamp for financial support through the FAPESP/CEPID Grant 2013/08293-7. B.I. acknowledges financial support from CNPq (process number \#153733/2024-1) and FAPESP (process number \#2024/11016-0). M.L.P.J. acknowledges financial support from the Federal District Research Support Foundation (FAPDF, grant 00193-00001807/2023-16), CNPq (grants 444921/2024-9 and 308222/2025-3), and CAPES (grant 88887.005164/2024-00).

\pagebreak

\begin{suppinfo}

\begin{figure}
\begin{center}
\includegraphics[width=0.5\linewidth]{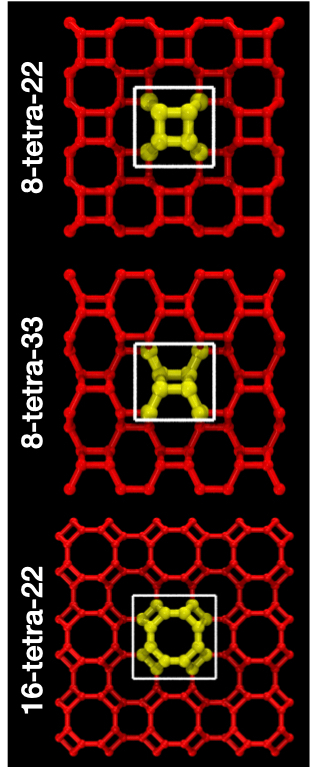}
\caption{Schematic representation (orthorhombic view) of tetragonal optimized structures 12-hexa-33, 24-hexa-20 and 36-hexa-33. The unit cell in yellow was replicated $2\times 2\times 2$.}
\label{fig:SI-1-tetra-UC}
\end{center}
\end{figure}

\begin{figure}
\begin{center}
\includegraphics[width=0.7\linewidth]{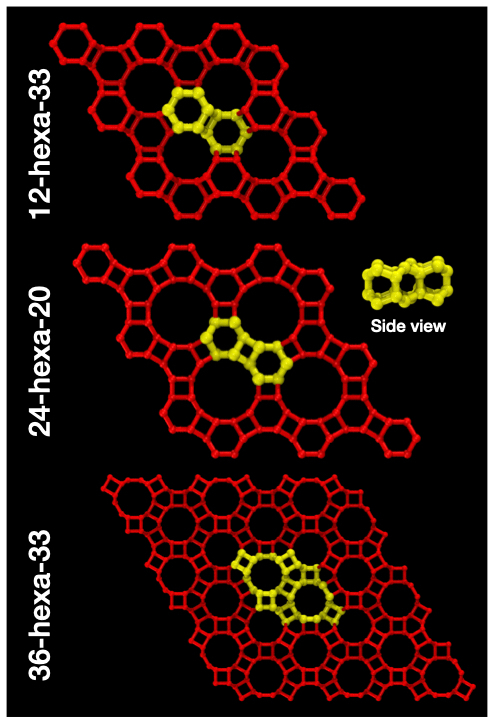}
\caption{Schematic representation (orthorhombic view) of hexagonal optimized structures 8-tetra-22, 8-tetra-33 and 16-tetra-22. The unit cell in yellow was replicated $2\times 2\times 2$.}
\label{fig:SI-2-hexa-UC}
\end{center}
\end{figure}

 \begin{figure}
\begin{center}
\includegraphics[width=0.8\linewidth]{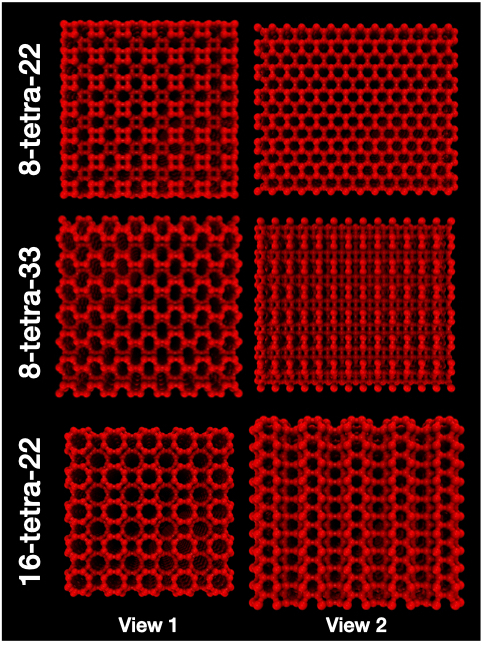}
\caption{Schematic representation of optimized unit cell replicated $3\times 3 \times 3$ for 8-tetra-22, 8-tetra-33 and 16-tetra-22 structures. }
\label{fig:SI-3-tetra}
\end{center}
\end{figure}

\begin{figure}
\begin{center}
\includegraphics[width=0.8\linewidth]{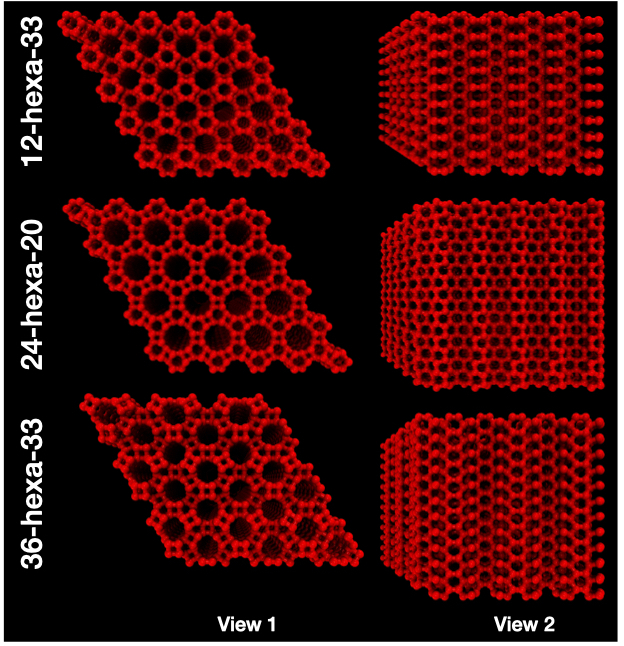}
\caption{Schematic representation of optimized unit cell replicated $3\times 3 \times 3$ for 12-hexa-33, 24-hexa-20 and 36-hexa-33 structures.}
\label{fig:SI-4-hexa}
\end{center}
\end{figure}

 \begin{figure}
\begin{center}
\includegraphics[width=0.8\linewidth]{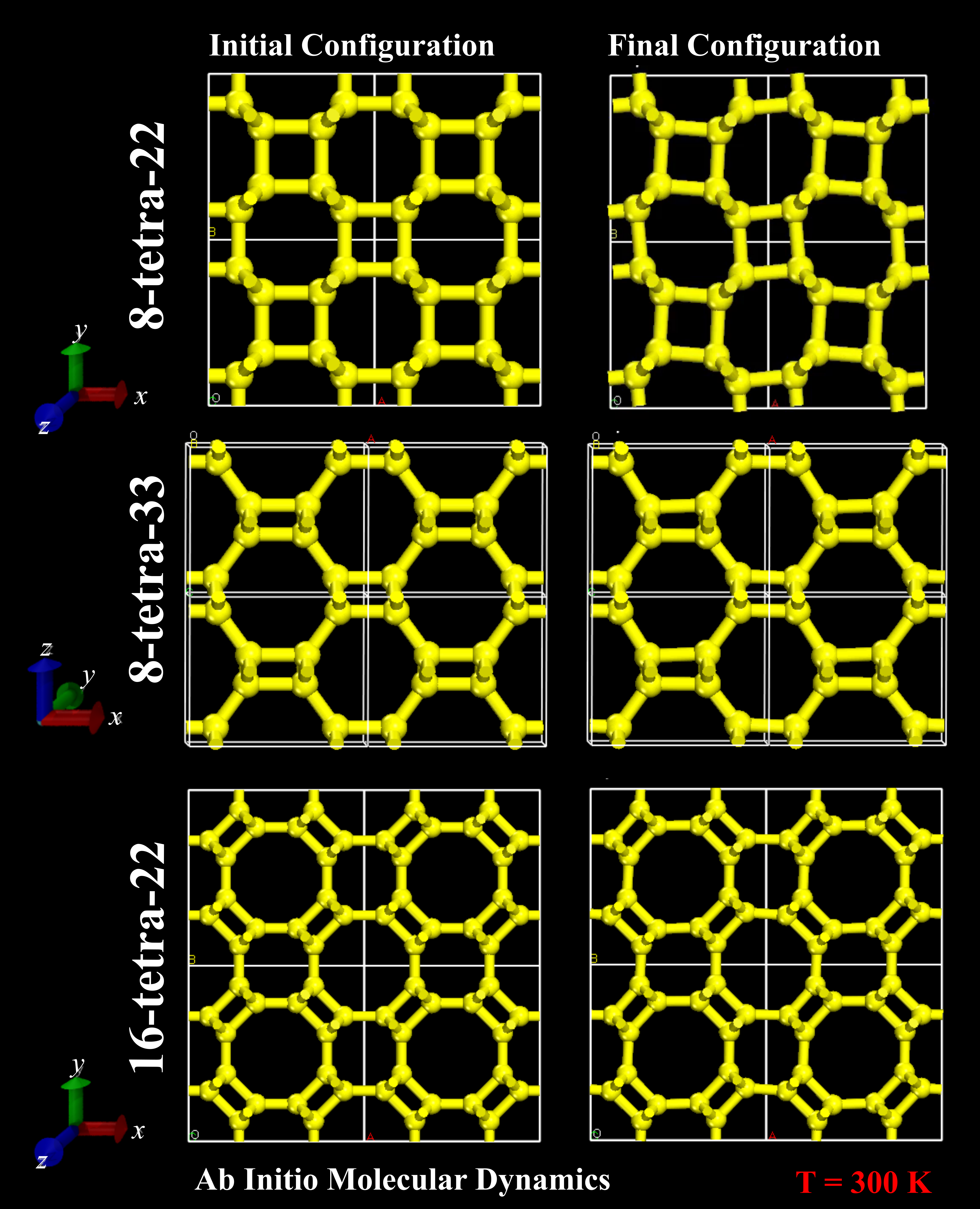}
\caption{Schematic representation of optimized unit cell replicated $3\times 3 \times 3$ for 12-hexa-33, 24-hexa-20 and 36-hexa-33 structures.}
\label{fig:SI-5-hexa-MD}
\end{center}
\end{figure}

 \begin{figure}
\begin{center}
\includegraphics[width=0.8\linewidth]{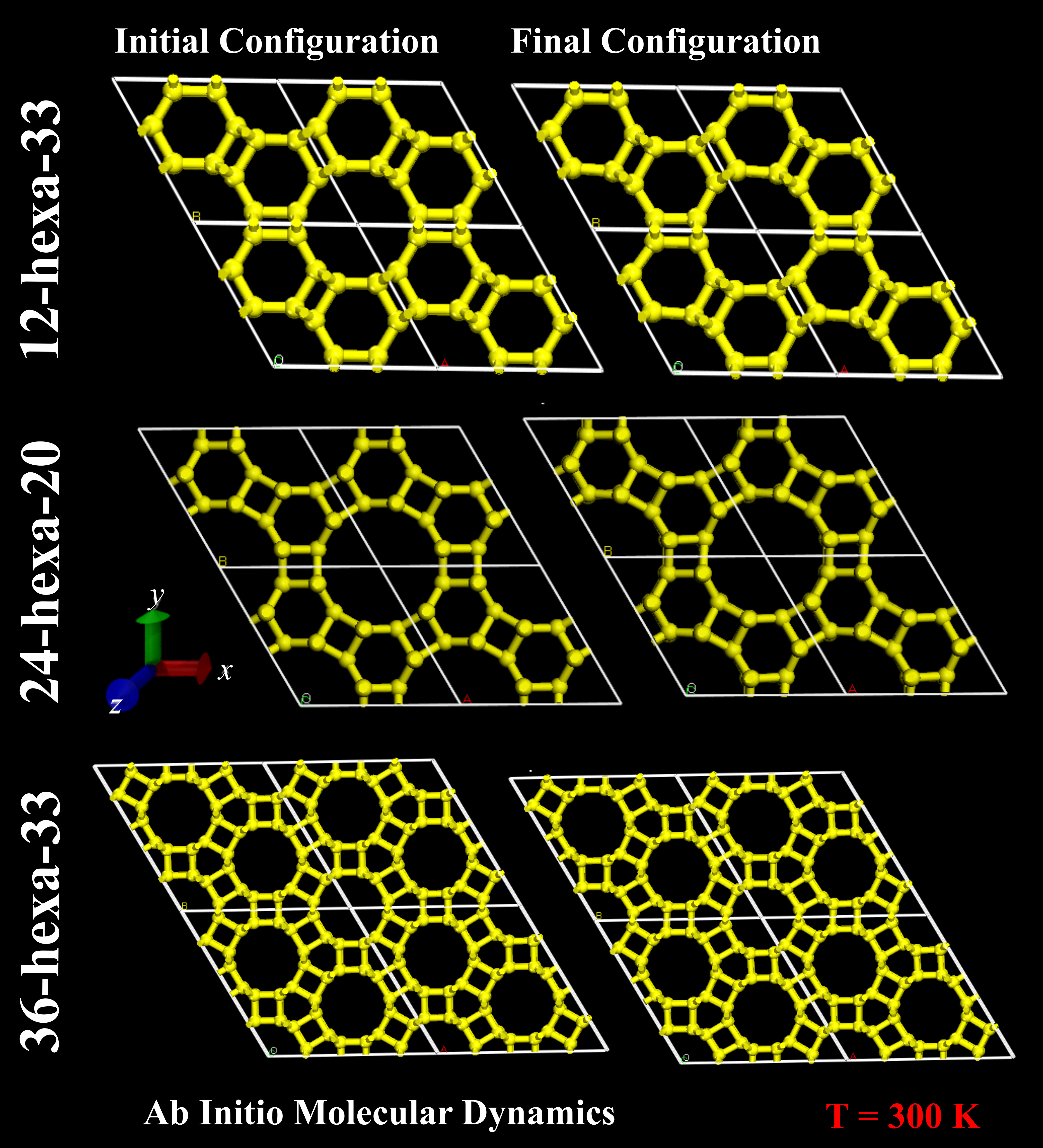}
\caption{Schematic representation of optimized unit cell replicated $3\times 3 \times 3$ for 8-tetra-22, 8-tetra-33 and 16-tetra-22 structures. }
\label{fig:SI-6-hexa-MD}
\end{center}
\end{figure}


\begin{figure}
\begin{center}
\includegraphics[width=0.83\linewidth]{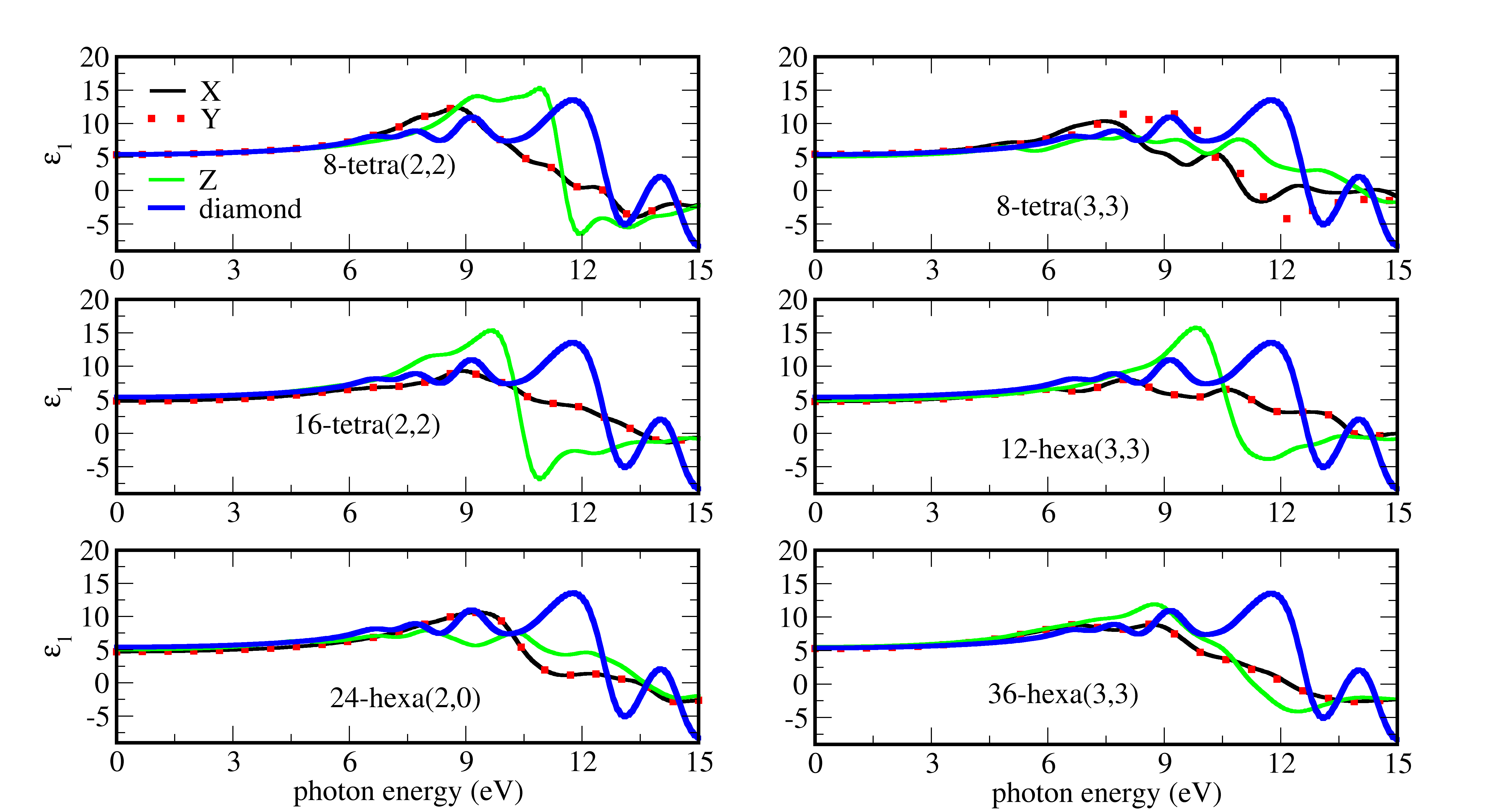}
\caption{Real part of the dielectric function {\it versus} photon energy,for an externally applies electric field polarized along $X,Y$ and $z$ directions} \label{fig:eps}
\end{center}
\end{figure}

\begin{figure}
\begin{center}
\includegraphics[width=0.83\linewidth]{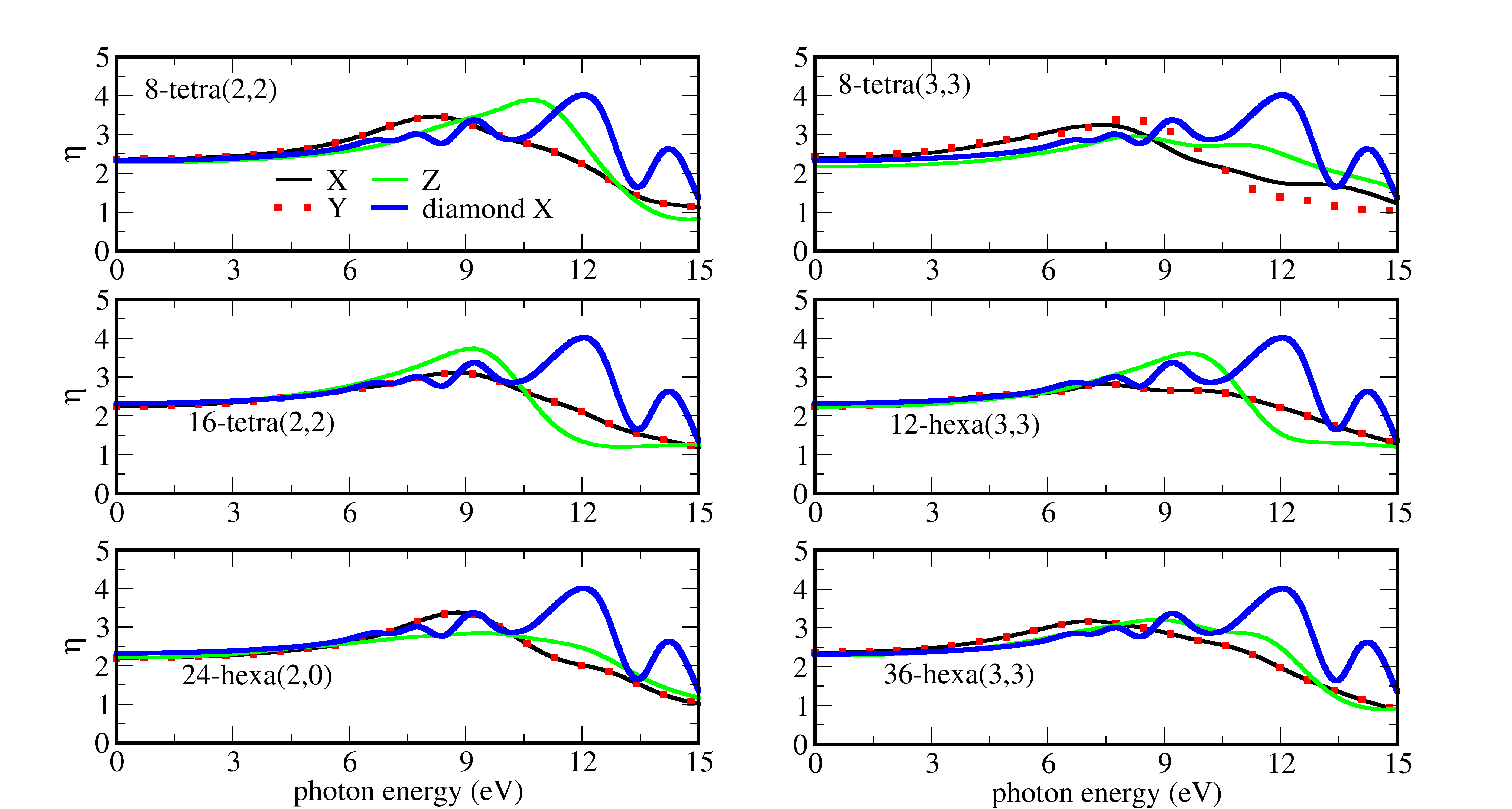}
\caption{Refractive index {\it versus} photon energy,for an externally applies electric field polarized along $X,Y$ and $z$ directions} \label{fig:refrac}
\end{center}
\end{figure}

\end{suppinfo}

\pagebreak
\bibliography{achemso-demo}

\end{document}